\documentclass[twocolumn,showpacs,preprintnumbers,amsmath,amssymb]{revtex4-2}
\usepackage{dsfont}
\usepackage{cases}
\usepackage{graphicx}
\usepackage{dcolumn}
\usepackage{bm}
\usepackage{amsmath}
\usepackage{appendix}

\usepackage[latin1]{inputenc}

\begin{document}

\title{Controlling Quantum Coherence of V-type Atom in Dissipative Cavity by Detuning and Weak Measurement Reversal}
\author{Qiying Pan\textsuperscript{1}}
\author{Fuhua Li\textsuperscript{1}}
\author{Hong-Mei Zou\textsuperscript{1,2}}
\email{zhmzc1997@hunnu.edu.cn}
\author{Zijin Liang\textsuperscript{1}}

\address{%
\textsuperscript{1}Key Laboratory of Low-Dimensional Quantum Structures and Quantum Control of Ministry of Education, Key Laboratory for Matter Microstructure and Function of Hunan Province, Department of Physics and Synergetic Innovation Center for Quantum Effects and Applications, Hunan Normal University, Changsha 410081, China}

\address{%
\textsuperscript{2}Hunan Research Center of the Basic Discipline for Quantum Effects and Quantum Technologies,
Hunan Normal University, Changsha 410081, China}

\date{\today}
\begin{abstract}
In this work, an interactive system composed of a V-type atom and a dissipative single-mode cavity is considered and the atomic quantum coherences are investigated under parameters including spontaneously generated interference (SGI), cavity-environment coupling, weak measurement and its reversal, and detuning between the atom and the cavity. The results indicate that strong coupling can induce coherence sudden death (CSD) and coherence sudden birth (CSB), while the non-zero SGI parameter only induces CSB. Detuning, however, may avoid both CSD and CSB. Moreover, detuning and weak measurement reversal can very effectively protect quantum coherence, while the SGI parameter, weak measurement, and strong coupling can accelerate its attenuation. The SGI parameter, detuning, weak measurement reversal, and strong coupling all promote the generation of coherence, whereas weak measurement alone can suppress it. In particular, the maximal coherent state can be very effectively protected and the coherent state can be prepared if all parameters are selected appropriately. Physical interpretations are also provided for these results.\par
\end{abstract}


\maketitle

\section{Introduction}
Quantum coherence, defined in terms of the off-diagonal elements of quantum states, is an essential building block of quantum mechanics and is closely related to the superposition principle \cite{Abdel-Khalek,Oumennana}. Quantum coherence has been considered as an important resource in quantum information, quantum key distribution, quantum biology and quantum optics \cite{Maccone,Chuang,Ma,Y. C. Cheng,Walls,Levi,Bera,Monda,Streltsov,Orszag,Hillery,Napoli,M. Piani}. However, quantum coherence is very fragile, and the inevitable coupling between quantum systems and their environments will lead to quantum decoherence in open systems \cite{Altowyan,Potnis,Yuce,Rahman,Algarni,Mortezapour}. Therefore, how to effectively protect and generate quantum coherence has always been a widely concerned and urgent issue.

Any actual physical system will suffer from unwanted interactions with external environments, thereby causing decoherence and destroying coherence. In recent years, open quantum systems have been widely studied from various aspects and by different methods \cite{Breuer,Alicki,H. M. Zou,H. M. Zou1,Rivas,Han,Brambilla,Weimer}. In theories of open quantum systems, environments are considered Markovian in the weak coupling regime but non Markovian in the strong coupling regime due to the memory and feedback effects of the environments. This non Markovian behavior is crucial for quantum information processing, as it has been realized in some experiments, particularly in the context of nanoplasmonic cavity quantum electrodynamics, which has enabled room-temperature strong coupling \cite{Almog,B.,Lei,Bundgaard-Nielsen,Chiang,W. M. Zhang,Tiwari,Jalali-Mola,Kokin,Lai,Do,Leng,Garcia}. The authors in Refs. \cite{Lostaglio,Behzadi,S. M. Fei,Ablimit,Yin,M. M.,Algarni1} investigated quantum coherence of open quantum systems and found that the quantum coherence will monotonically reduce to zero in Markovian environments but it can be effectively protected in non Markovian environments. The result in \cite{H. M. Zou2} showed that the non Markovianity can act as a protector of qubit coherence. Bromley et al. \cite{Bromley} analyzed the freezing conditions of quantum coherence in an open quantum system. Mani et al. \cite{Mani} investigated the cohering and decohering power of various decoherence channels. Besides, quantum coherence can also be effectively preserved in open quantum systems by increasing the detuning and reasonably utilizing environmental noises \cite{Shahri,Xu1,Lim,Youssry}.

On the other hand, various controlling schemes have also been put forward to eliminate decoherence and protect quantum coherence. For example, the decoherence process can be monitored by quantum feedback control, which may mitigate the loss of coherence \cite{J. Zhang,Geremia}. Besides, there are several other approaches, such as external classical driving \cite{Rahman1,Gholipour}, quantum Zeno effect \cite{Maniscalco,Kondo}, decoherence-free subspace \cite{Kwiat,Qin}, dynamical decoupling sequences \cite{Viola, R. B. Liu,G. S. Uhrig,Pryadko}, entanglement distillation \cite{Kwiat1,Pan,Dong}, and weak measurement and its reversal \cite{H. M. Zou2,Q. Wang,Kim,J. He,J. He1,J. He2,Zhang,J. He3}.

In the above-mentioned researches, most focus on qubit systems. However, it is very worthwhile to pour more efforts into studying open multilevel quantum systems because they are superior physical resources than qubits in quantum information processing such as quantum cryptography \cite{Bru?,Cerf} and quantum interference \cite{Scully}. Recently, certain techniques for controlling coherence in open qubits have also been extended to qutrits, for example, Wang and Li et al. studied the entanglement dynamics of two V-type atoms in a common dissipative cavity \cite{J. Wang,J. Wang1}. Quantum interference plays a positive role in protecting quantum coherence in qutrit systems \cite{Zeng}. In a three-level quantum system, the optimal coherence survival can be facilitated by properly adjusting the Gaussian noisy parameters \cite{Zangi}. Moreover, weak measurement and its reversal can effectively suppress amplitude-damping decoherence in a qutrit system \cite{C. Yao,Zou}, and the sensitivity of detecting decoherence can be effectively amplified by quantum control pulse sequences \cite{Xu}. The results in \cite{Faizi} indicate that quantum coherence in an open three-level atom can be protected using auxiliary atoms. Decoherence can be effectively suppressed in an accelerated qubit-qutrit system \cite{M. Y.}.

Motivated by these works, we will investigate the quantum coherence dynamics of a V-type atom in a dissipative cavity under detuning and weak measurement reversal. We aim to explore whether the coherence of a qutrit system can be maintained and produced by the atom-cavity detuning and the weak measurement reversal. The results demonstrate that  the detuning and weak measurement reversal can very effectively protect quantum coherence, while the SGI parameter, weak measurement, and strong coupling can accelerate its attenuation. The SGI parameter, detuning, weak measurement reversal, and strong coupling all promote coherence generation, whereas weak measurement alone can suppress it.

The outline is following. In Section II, we introduce a physical model and obtain its analytical solution. In Section III, we introduce weak measurement and its reversal. In Section IV, we analytically determine quantum coherence of a V-type atom system. Results and discussions are provided in Section V. The paper is ended with a brief conclusion in Section VI.
\begin{figure}[tbp]
	\includegraphics[width=9cm,height=5cm]{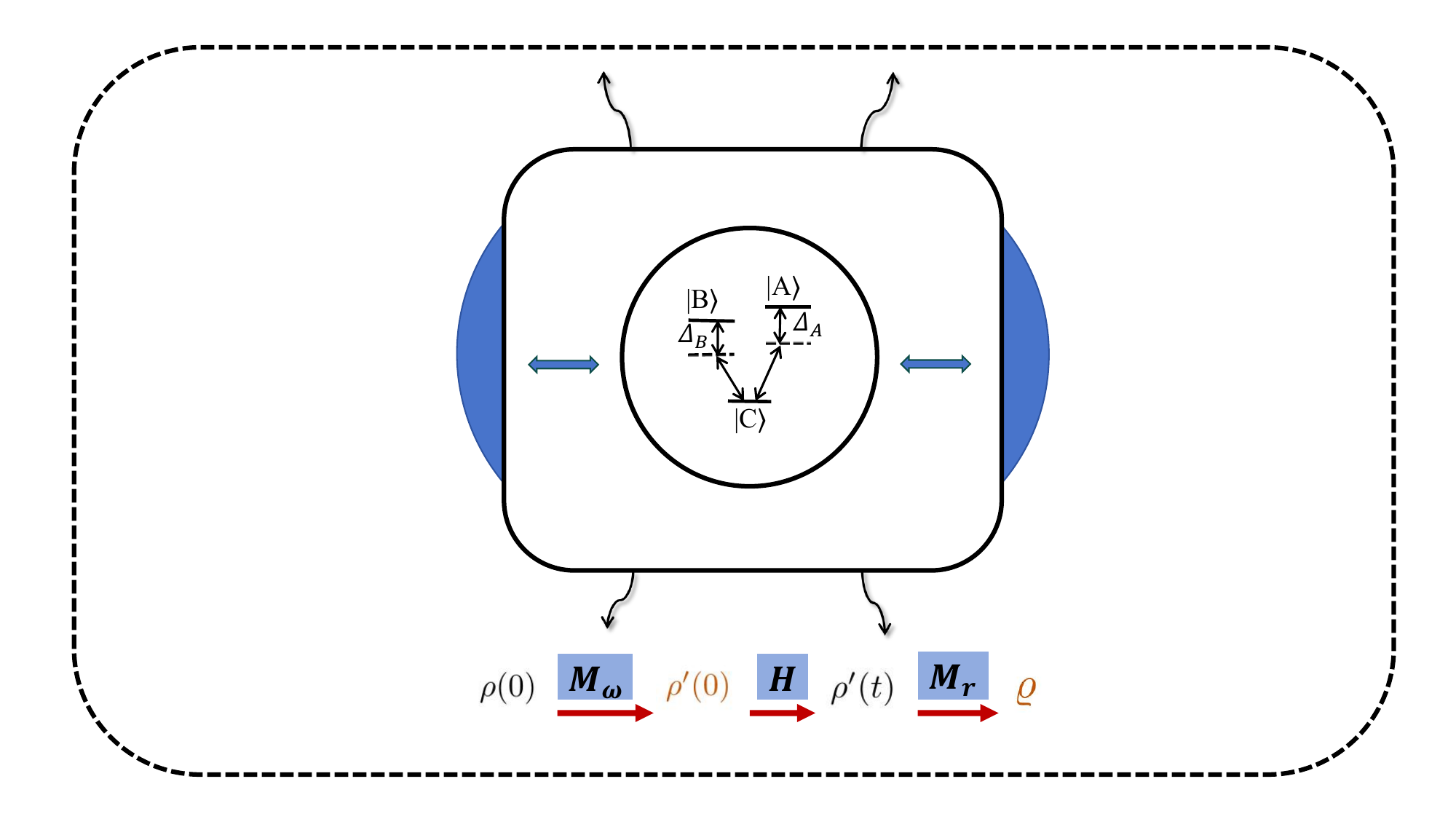}
	\caption{A V-type qutrit atomic system interacts with a dissipative cavity coupled to an external environment, and the atom is submitted to the weak measurement and the weak measurement reversal. Meanwhile, the atomic transition frequency is detuned from the cavity eigenfrequency.}
	\label{figure:1}
\end{figure}

\section{Physical model}
In Refs. \cite{J. Wang,J. Wang1}, we studied the system of two V-type atoms in a common dissipative cavity using the Fano theorem and the Schr\"{o}dinger equation. That is to say, by assuming that the atomic transition frequency resonates with both the cavity eigenfrequency and the environmental center frequency, we focused on how the entanglement dynamics depend on the dipole-dipole interaction between two atoms \cite{J. Wang}. In Ref. \cite{J. Wang1}, we studied the influence of weak measurement reversal on entanglement dynamics. We assumed that the atomic transition frequency resonates with the cavity eigenfrequency but is detuned from the environmental center frequency.

However, in this work, we only consider a V-type atom interacting with a dissipative cavity and assume that the atomic transition frequency is detuned from the cavity eigenfrequency but resonates with the environmental center frequency.

As shown in FIG. 1, the atom has two upper excited states $|A\rangle$, $|B\rangle$ which can spontaneously decay into ground state $|C\rangle$ with frequencies $\omega _{A}$, $\omega _{B}$, and $\omega _{C}$ respectively. This atom interacts with a dissipative cavity which couples with an external environment described by a series of continuum harmonic oscillators. In the rotating-wave approximation \cite{G. S.} and with $\hbar=1$, the free Hamiltonian composed of the atom and the dissipative cavity is
\begin{equation} \label{Eq1}
\begin{split}
\hat{H}_{0}&=\hat{H}_{atom}+\hat{H}_{cavity} \\
&=\omega_{A}|A\rangle\langle A|+\omega_{B}| B\rangle\langle B|+\omega_{C}| C\rangle\langle C|\\
&+\omega_{0} \hat{a}^{\dagger} \hat{a}+\int_{0}^{\infty} \eta \hat{b}^{\dagger}(\eta) \hat{b}(\eta) d \eta \\
&+\int_{0}^{\infty}\left\{g(\eta) \hat{a}^{\dagger} \hat{b}(\eta)+H . C .\right\} d \eta
\end{split}
\end{equation}%
where the first three terms are the atomic Hamiltonian and the last three terms are the Hamiltonian of the dissipative cavity. $\omega _{0}$ is the cavity eigenfrequency and $\hat{a}$ ($\hat{a}^{\dag}$) is the annihilation (creation) operator of the cavity. $g(\eta)=\sqrt{\kappa / \pi}$ is the coupling coefficient of cavity-environment and $\kappa$ is the spectral width of coupling. $\hat{b}^{\dag}(\eta)$ ($\hat{b}(\eta)$) is the creation (annihilation) operator of environment in the $\eta$-th mode. $H.C.$ is the Hermitian conjugation \cite{J. Wang,J. Wang1}. The interaction Hamiltonian between the atom and the cavity is
\begin{equation}\label{Eq2}
\begin{split}
\hat{H}_{I} & =g_{A} \hat{a}|A\rangle\langle C|+g_{A}^{*} \hat{a}^{\dagger}| C\rangle\langle A| \\
& +g_{B} \hat{a}|B\rangle\langle C|+g_{B}^{*} \hat{a}^{\dagger}| C\rangle\langle B|
\end{split}
\end{equation}%
where $g_{A}$ ($g_{B}$) represents the coupling strength between the atomic state $|A\rangle$ ($|B\rangle$) and the cavity mode.

According to the Fano theorem, the Hamiltonian of the dissipative cavity can be diagonalized as  $\hat{H}_{cavity}=\int \omega \hat{T}^{\dagger}(\omega) \hat{T}(\omega) d \omega$ where $\hat{T}(\omega)=\alpha(\omega) \hat{a}+\int \beta(\omega, \eta)\hat{b}(\eta) d \eta$ and  $\alpha(\omega)=\frac{\sqrt{\kappa / \pi}}{\omega-\omega_{0}+i \kappa}$  \cite{J. Wang,J. Wang1,Nourmandipour}. Then, the free Hamiltonian (Eq.~(\ref{Eq1})) and the interaction Hamiltonian (Eq.~(\ref{Eq2})) become respectively as
\begin{equation}\label{Eq3}
\begin{split}
\hat{H}_{0} & =\omega_{A}|A\rangle\langle A|+\omega_{B}| B\rangle\langle B|+\omega_{C}| C\rangle\langle C| \\
& +\int \omega \hat{T}^{\dagger}(\omega) \hat{T}(\omega) d \omega
\end{split}
\end{equation}%
and
\begin{equation} \label{Eq4}
\begin{split}
\hat{H}_{I} & =\int\left(g_{A} \alpha^{*}(\omega) \hat{T}(\omega)|A\rangle\langle C|+H . C .\right) d \omega \\
& +\int\left(g_{B} \alpha^{*}(\omega) \hat{T}(\omega)|B\rangle\langle C|+H . C .\right) d \omega
\end{split}
\end{equation}%

Let the initial and time-evolved states of the total system be respectively given by
\begin{equation} \label{Eq6}
|\psi(0)\rangle =\left(D_{A}(0)|A\rangle+D_{B}(0)|B\rangle+D_{C}(0)|C\rangle\right)_{S} \otimes|0\rangle_{E}
\end{equation}%
with $\left|D_{A}(0)\right|^{2}+\left|D_{B}(0)\right|^{2}+\left|D_{C}(0)\right|^{2}=1$, and
\begin{equation} \label{Eq7}
\begin{split}
|\psi(t)\rangle & =\left(D_{A}(t)|A\rangle+D_{B}(t)|B\rangle+D_{C}(t)|C\rangle\right)_{S} \otimes|0\rangle_{E} \\
& +\int D_{\omega}(t)|C\rangle_{S}\left|1_{\omega}\right\rangle_{E} d \omega
\end{split}
\end{equation}%
with $\left|D_{A}(t)\right|^{2}+\left|D_{B}(t)\right|^{2}+\left|D_{C}(t)\right|^{2}+\left|D_{\omega}(t)\right|^{2}=1$, where
$|0\rangle_{E}$ means that the reservoir is in the vacuum state and $|1_{\omega}\rangle_{E}$ is the reservoir state having one excitation merely in the $\omega$-th mode. Substituting Eqs.~(\ref{Eq3})-(\ref{Eq7}) into the Schr\"{o}dinger equation in the interaction picture \cite{J. Wang,J. Wang1}
\begin{equation}\label{Eq5}
i \frac{d}{d t}|\psi(t)\rangle=\hat{H}_{i n t}(t)|\psi(t)\rangle
\end{equation}%
with $\hat{H}_{int}(t)  =e^{i \hat{H}_{0} t} \hat{H}_{I} e^{-i \hat{H}_{0} t}$, we can get $D_{C}(t)=D_{C}(0)$ and
\begin{equation} \label{Eq8}
\frac{d D_{m}(t)}{d t}=-\sum_{n=A, B} \int_{0}^{t} f_{m n}\left(t-t^{\prime}\right) D_{n}\left(t^{\prime}\right) d t^{\prime} \quad m=A, B
\end{equation}%
where the kernel
\begin{equation} \label{Eq9}
f_{m n}\left(t-t^{\prime}\right)=\int J_{m n}(\omega) e^{i\left(\omega_{m}-\omega_{C}-\omega\right) t-i\left(\omega_{n}-\omega_{C}-\omega\right) t^{\prime}} d \omega
\end{equation}%

If the spectral density $J_{m n}(\omega)$ of environment has the Lorentzian form and the cavity eigenfrequency resonates with the central frequency of the external environment, i.e.
\begin{equation} \label{Eq10}
J_{m n}(\omega)=\frac{1}{2 \pi} \frac{\gamma_{m n} \kappa^{2}}{\left(\omega-\omega_{0}\right)^{2}+\kappa^{2}}
\end{equation}%
where $\gamma_{m n}=\frac{2 g_{m} g_{n}^{*}}{\kappa}$ ($m, n=A, B$) is the relaxation rate of the two upper excited states, $\gamma_{mm}=\gamma_{m}$ and
$\gamma_{m n}=\sqrt{\gamma_{m} \gamma_{n}} \theta$ $( m \neq n,|\theta| \leq 1)$
where $\theta$ is the SGI parameter between two decay channels $|A\rangle\rightarrow|C\rangle$ and $|B\rangle\rightarrow|C\rangle$, which relies on the angle between two transition dipole moments. $\theta=0$ indicates that the dipole moments of the transitions are perpendicular to each other. Instead, $\theta=1$ means the two dipole moments are parallel. Let the two excited states be degenerate, i.e. $\omega_{A}=\omega_{B}$ and $\gamma_{A B}=\gamma_{B A}=\gamma_{0} \theta$, which $\gamma_{0}$ is the decay coefficient of atomic excited states. $\kappa\gg2\gamma_{0}$ indicates the weak (Markovian) coupling between the cavity and its environment while $\kappa\ll2\gamma_{0}$ means the strong (non Markovian) coupling \cite{Bellomo}. Also, assume the detuning between the atomic transition frequency and the cavity eigenfrequency is $\Delta$, i.e. $\Delta=\omega_{0}-\left(\omega_{m}-\omega_{C}\right)$ $(m=A,B)$. Then, the kernels (Eq.~(\ref{Eq9})) can be expressed as
\begin{equation} \label{Eq11}
\begin{split}
f_{A A}\left(t-t^{\prime}\right) & =f_{B B}\left(t-t^{\prime}\right)=f\left(t-t^{\prime}\right) \\
& =\int_{0}^{\infty} d \omega J(\omega) e^{-i \Delta\left(t-t^{\prime}\right)} e^{i\left(\omega_{0}-\omega\right)\left(t-t^{\prime}\right)}\\
f_{A B}\left(t-t^{\prime}\right) & =f_{B A}\left(t-t^{\prime}\right)=f^{\prime}\left(t-t^{\prime}\right) \\
& =\int_{0}^{\infty} d \omega J^{\prime}(\omega) e^{-i \Delta\left(t-t^{\prime}\right)} e^{i\left(\omega_{0}-\omega\right)\left(t-t^{\prime}\right)}
\end{split}
\end{equation}%
where $J^{\prime}(\omega)=\theta J(\omega)$.
Taking Eq.~(\ref{Eq10}) into Eq.~(\ref{Eq11}), we can obtain
\begin{equation} \label{Eq12}
\begin{split}
f\left(t-t^{\prime}\right) & =\frac{\gamma_{0} \kappa}{2} e^{(-\kappa+i \Delta)\left(t-t^{\prime}\right)} \\
f^{\prime}\left(t-t^{\prime}\right) & =\frac{\gamma_{0} \theta \kappa}{2} e^{(-\kappa+i \Delta)\left(t-t^{\prime}\right)}
\end{split}
\end{equation}%

Through the Laplace transform and its inverse transform \cite{J. Wang,J. Wang1}, we obtain the amplitudes consequently
\begin{equation} \label{Eq13}
\begin{split}
D_{A}(t)&=\frac{D^{+}(t)+D^{-}(t)}{2} \\
D_{B}(t)&=\frac{D^{+}(t)-D^{-}(t)}{2} \\
D_{C}(t)&=D_{C}(0)
\end{split}
\end{equation}%
here
\begin{equation}  \label{Eq14}
\begin{split}
D^{ \pm}(t)=\mathcal{G}^{ \pm}(t) D^{ \pm}(0)
\end{split}
\end{equation}%
where
\begin{equation} \label{Eq15}
\begin{split}
D^{ \pm}(0)&=D_{A}(0) \pm D_{B}(0)\\
\mathcal{G}^{ \pm}(t)&=e^{-(\kappa+i \Delta) t / 2}\\
&\times\left\{\cosh \left(\frac{R^{ \pm} t}{2}\right)+\frac{\kappa+i \Delta}{R^{ \pm}} \sinh \left(\frac{R^{ \pm} t}{2}\right)\right\}
\end{split}
\end{equation}%
and
\begin{equation}  \label{Eq16}
R^{ \pm}=\sqrt{(\kappa+i \Delta)^{2}-2 \gamma_{0}(1 \pm \theta) \kappa}
\end{equation}%
From  Eqs.~(\ref{Eq13})-(\ref{Eq16}), we know the analytical expressions of probability amplitudes
\begin{equation}\label{Eq17}
\begin{aligned}
& D_{A}(t)=\mathcal{Q}_{1}(t) D_{A}(0)+\mathcal{Q}_{2}(t) D_{B}(0) \\
& D_{B}(t)=\mathcal{Q}_{2}(t) D_{A}(0)+\mathcal{Q}_{1}(t) D_{B}(0) \\
& D_{C}(t)=D_{C}(0)
\end{aligned}
\end{equation}
where
\begin{equation}\label{Eq18}
\begin{split}
\mathcal{Q}_{1}(t)=\frac{\mathcal{G}^{+}(t)+\mathcal{G}^{-}(t)}{2} \\
\mathcal{Q}_{2}(t)=\frac{\mathcal{G}^{+}(t)-\mathcal{G}^{-}(t)}{2}
\end{split}
\end{equation}%

We can gain the atomic density operator in the basis $\{|C\rangle,|B\rangle,|A\rangle\}$ from Eq.~(\ref{Eq7}) by tracing the freedom degree of the reservoir
\begin{equation}\label{Eq20}
\begin{aligned}
\rho(t)=
& \left(\begin{array}{ccc}
|D_{\omega}(t)|^{2}+|D_{C}(t)|^{2} & D_{C}(t) D_{B}^*(t) & D_{C}(t) D_{A}^*(t) \\
D_{B}(t) D_{C}^*(t) & |D_{B}(t)|^2 & D_{B}(t) D_{A}^*(t) \\
D_{A}(t) D_{C}^*(t) & D_{A}(t) D_{B}^*(t) & |D_{A}(t)|^2
\end{array}\right)
\end{aligned}
\end{equation}

\section{weak measurement and weak measurement reversal}
For the single V-type atom system, the operators of the weak measurement and its reversal can be written as \cite{J. Wang1}
\begin{equation}\label{Eq21}
M^{w}_{p,q}=\left(\begin{array}{ccc}
1 & 0 & 0 \\
0 & \sqrt{1-p} & 0 \\
0 & 0 & \sqrt{1-q}
\end{array}\right)
\end{equation}%
and
\begin{equation}\label{Eq22}
M^{r}_{p_{r},q_{r}}=\left(\begin{array}{ccc}
\sqrt{\left(1-p_{r}\right)\left(1-q_{r}\right)} & 0 & 0 \\
0 & \sqrt{1-q_{r}} & 0 \\
0 & 0 & \sqrt{1-p_{r}}
\end{array}\right)
\end{equation}%
where $p$($q$) and $p_{r}$($q_{r}$) represent the weak and reversing measurement strengths, satisfying $0\leq p(q)<1$ and $0\leq p_{r}(q_{r}<1$), respectively. Only considering the cases $p=q$ and $p_{r}=q_{r}$, the measurement operators are simplified as
\begin{equation}\label{Eq23}
M^w=\left(\begin{array}{ccc}
1 & 0 & 0 \\
0 & \sqrt{1-p} & 0 \\
0 & 0 & \sqrt{1-p}
\end{array}\right)
\end{equation}
\begin{equation}\label{Eq24}
M^r=\left(\begin{array}{ccc}
1-p_r & 0 & 0 \\
0 & \sqrt{1-p_r} & 0 \\
0 & 0 & \sqrt{1-p_r}
\end{array}\right)
\end{equation}

The atomic initial density operator is
\begin{equation}\label{Eq25}
\begin{aligned}
\rho(0)=
\left(\begin{array}{ccc}
|D_{C}(0)|^2 & D_{C}(0) D_{B}^*(0) & D_{C}(0) D_{A}^*(0) \\
D_{B}(0) D_{C}^*(0) & |D_{B}(0)|^2 & D_{B}(0) D_{A}^*(0) \\
D_{A}(0) D_{C}^*(0) & D_{A}(0) D_{B}^*(0) & |D_{A}(0)|^2
\end{array}\right)
\end{aligned}
\end{equation}
which is subjected to a prior weak measurement and a post weak measurement reversal. After the weak measurement, Eq.~(\ref{Eq25}) becomes as $\rho^{\prime}(0)=M^\omega \rho(0) M^{\omega\dag}$, the elements of $\rho^{\prime}(0)$ are
\begin{equation}\label{Eq26}
\begin{array}{l}
\rho_{11}^{\prime}(0)=\left|D_{C}(0)\right|^{2} \\
\rho_{12}^{\prime}(0)=D_{C}(0) D_{B}^{*}(0)(1-p)^{\frac{1}{2}} \\
\rho_{13}^{\prime}(0)=D_{C}(0) D_{A}^{*}(0)(1-p)^{\frac{1}{2}} \\
\rho_{21}^{\prime}(0)=D_{B}(0) D_{C}^{*}(0)(1-p)^{\frac{1}{2}} \\
\rho_{22}^{\prime}(0)=\left|D_{B}(0)\right|^{2}(1-p) \\
\rho_{23}^{\prime}(0)=D_{B}(0) D_{A}^{*}(0)(1-p) \\
\rho_{31}^{\prime}(0)=D_{A}(0) D_{C}^{*}(0)(1-p)^{\frac{1}{2}} \\
\rho_{32}^{\prime}(0)=D_{A}(0) D_{B}^{*}(0)(1-p) \\
\rho_{33}^{\prime}(0)=|D_{A}(0)|^{2}(1-p)
\end{array}
\end{equation}
here $D_A^{\prime}(0)=D_A(0)(1-p)^{\frac{1}{2}}$, $D_B^{\prime}(0)=D_B(0)(1-p)^{\frac{1}{2}}$ and $D_C^{\prime}(0)=D_C(0)$. Substituting $D_A^{\prime}(0)$, $D_B^{\prime}(0)$ and $D_C^{\prime}(0)$ into Eq.~(\ref{Eq17}), we can obtain the probability amplitudes under the Hamiltonian as
\begin{equation}\label{Eq27}
\begin{aligned}
& D_{A}^{\prime}(t)=\mathcal{Q}_{1}(t) D_{A}^{\prime}(0)+\mathcal{Q}_{2}(t) D_{B}^{\prime}(0) \\
& D_{B}^{\prime}(t)=\mathcal{Q}_{2}(t) D_{A}^{\prime}(0)+\mathcal{Q}_{1}(t) D_{B}^{\prime}(0) \\
& D_{C}^{\prime}(t)=D_C^{\prime}(0)
\end{aligned}
\end{equation}
$\rho^{\prime}(t)$ can be written as
\begin{equation}\label{Eq28}
\begin{aligned}
\rho^{\prime}(t)=
& \left(\begin{array}{ccc}
|D_{\omega}^{\prime}(t)|^{2}+|D_{C}^{\prime}(t)|^{2} & D_{C}^{\prime}(t) D_{B}^{* \prime}(t) & D_{C}^{\prime}(t) D_{A}^{* \prime}(t) \\
D_{B}^{\prime}(t) D_{C}^{* \prime}(t) & |D_{B}^{\prime}(t)|^2 & D_{B}^{\prime}(t) D_{A}^{* \prime}(t) \\
D_{A}^{\prime}(t) D_{C}^{* \prime}(t) & D_{A}^{\prime}(t) D_{B}^{* \prime}(t) & |D_{A}^{\prime}(t)|^2
\end{array}\right)
\end{aligned}
\end{equation}
where $|D_{\omega}^{\prime}(t)|^{2}=1-|D_{A}^{\prime}(t)|^2-|D_{B}^{\prime}(t)|^2-|D_{C}^{\prime}(t)|^2$.

Next, we perform the weak measurement reversal on $\rho^{\prime}(t)$, i.e. $\rho=M^r \rho^{\prime}(t) M^{r\dag}$. As the measurement operators in Eqs.~(\ref{Eq23})-(\ref{Eq24}) are non-unitary, we need to normalize the density operator $\rho$ using the normalization factor $C_1=(\left|D_\omega^{\prime}(t)\right|^2+\left|D_{C}^{\prime}(t))\right|^2)\left(1-p_r\right)^2+\left|D_A^{\prime}(t)\right|^2\left(1-p_r\right)+\left|D_B^{\prime}(t)\right|^2\left(1-p_r\right)$, the elements of $\rho$ are
\begin{equation}\label{Eq29}
\begin{array}{l}
\rho_{11}=\frac{1}{C_{1}}\left(\left|D_{\omega}^{\prime}(t)\right|^{2}+\left|D_{C}^{\prime}(t)\right|^{2}\right)\left(1-p_{r}\right)^{2} \\
\rho_{12}=\frac{1}{C_{1}} D_{C}^{\prime}(t) D_{B}^{* \prime}(t)\left(1-p_{r}\right)^{\frac{3}{2}} \\
\rho_{13}=\frac{1}{C_{1}} D_{C}^{\prime}(t) D_{A}^{* \prime}(t)\left(1-p_{r}\right)^{\frac{3}{2}} \\
\rho_{21}=\frac{1}{C_{1}} D_{B}^{\prime}(t) D_{C}^{* \prime}(t)\left(1-p_{r}\right)^{\frac{3}{2}} \\
\rho_{22}=\frac{1}{C_{1}}\left|D_{B}^{\prime}(t)\right|^{2}\left(1-p_{r}\right) \\
\rho_{23}=\frac{1}{C_{1}} D_{B}^{\prime}(t) D_{A}^{* \prime}(t)\left(1-p_{r}\right) \\
\rho_{31}=\frac{1}{C_{1}} D_{A}^{\prime}(t) D_{C}^{* \prime}(t)\left(1-p_{r}\right)^{\frac{3}{2}} \\
\rho_{32}=\frac{1}{C_{1}} D_{A}^{\prime}(t) D_{B}^{* \prime}(t)\left(1-p_{r}\right) \\
\rho_{33}=\frac{1}{C_{1}}\left|D_{A}^{\prime}(t)\right|^{2}\left(1-p_{r}\right)
\end{array}
\end{equation}

\section{quantum coherence of single V-type atom system}
Quantum coherence is ascribed to the off-diagonal elements of the density matrix and strictly depends on the chosen basis. Among a number of quantum coherence measures, the $\ell_{1}$ norm of coherence is an intuitive and widely used measure. Here, we also use the $\ell_{1}$ norm to quantify the atomic quantum coherence, which is defined as \cite{Baumgratz,C. S. Yu}.
\begin{equation}\label{Eq30}
\xi(\rho)=\sum_{i, j(i \neq j)}\left|\rho_{i j}\right|
\end{equation}%
where $\left|\rho_{i j}\right|$ is the absolute value of the off-diagonal elements in Eq.~(\ref{Eq29}). Finally, the coherence of the V-type atom system can be analytically expressed as
\begin{equation}\label{Eq31}
\begin{aligned}
\xi(\rho) &= \frac{1}{C_{1}} \\
&\left( \left|D_{C}^{\prime}(t) D_{B}^{* \prime}(t)\left(1-p_{r}\right)^{\frac{3}{2}}\right|
+ \left|D_{C}^{\prime}(t) D_{A}^{* \prime}(t)\left(1-p_{r}\right)^{\frac{3}{2}}\right| \right. \\
&+ \left|D_{B}^{\prime}(t) D_{C}^{* \prime}(t)\left(1-p_{r}\right)^{\frac{3}{2}}\right|
+ \left|D_{B}^{\prime}(t) D_{A}^{* \prime}(t)\left(1-p_{r}\right)\right| \\
&\left. + \left|D_{A}^{\prime}(t) D_{C}^{* \prime}(t)\left(1-p_{r}\right)^{\frac{3}{2}}\right|
+ \left|D_{A}^{\prime}(t) D_{B}^{* \prime}(t)\left(1-p_{r}\right)\right| \right)
\end{aligned}
\end{equation}%

\section{Results and Discussions}
In this subsection. we will discuss quantum coherent dynamics in detail in three different cases, such as the initial state phase, the atom-cavity resonance and the atom-cavity detuning.

\subsection{Initial state phase}
Firstly, we analyze that the quantum coherence depends on the phases of the initial state $|\psi(0)\rangle$.
Assume that the probability amplitudes in Eq.~(\ref{Eq6}) are respectively as $D_{A}(0)=\cos\alpha$, $D_{B}(0)=e^{i\beta}\sin\alpha$ and $D_{C}(0)=0$. Namely,
\begin{equation} \label{Eq32}
|\psi(0)\rangle =\left(\cos\alpha|A\rangle+e^{i\beta}\sin\alpha|B\rangle\right)_{S} \otimes|0\rangle_{E}
\end{equation}%
In FIG. 2, we provide the dependence of coherence dynamics on the initial state phases in the strong coupling regime ($\gamma_0/\kappa=10$). FIG. 2(a) shows the coherence dynamics as a function of $\beta$ and $t$ when $\alpha=\frac{\pi}{4}$. From FIG. 2(a), we can see that the quantum coherence is obviously dependent on $t$ but independent of $\beta$. For a certain value of $\beta$, the quantum coherence oscillates damply when time $t$ increases. That is to say, the quantum coherence first rapidly decays from 1.0 to zero and then again revives to a certain value in a very short time. For different $\beta$, the quantum coherence always oscillates damply in the same rule as time $t$ increases. FIG. 2(b) depicts the coherence dynamics as a function of $\alpha$ and $t$ when $\beta=0$. From FIG. 2(b), we find that, the quantum coherence is not only dependent on $t$ but also dependent on $\alpha$. For a certain time $t$, the periodic oscillation will occur in the quantum coherence as $\alpha$ increases. For a certain value of $\alpha$, the quantum coherence oscillates damply when time $t$ increases. For different values of $\alpha$, the quantum coherence exhibits diverse evolution curves. For example, the quantum coherence is always zero when $\alpha=\frac{\pi}{2}$, while it decays from 1.0 and then oscillates when $\alpha=\frac{\pi}{4}$.

To better understand how other parameters affect the quantum coherence, we will set $\beta=0$ in our subsequent discussions.

\begin{figure}[tbp]
	\includegraphics[width=4cm,height=3cm]{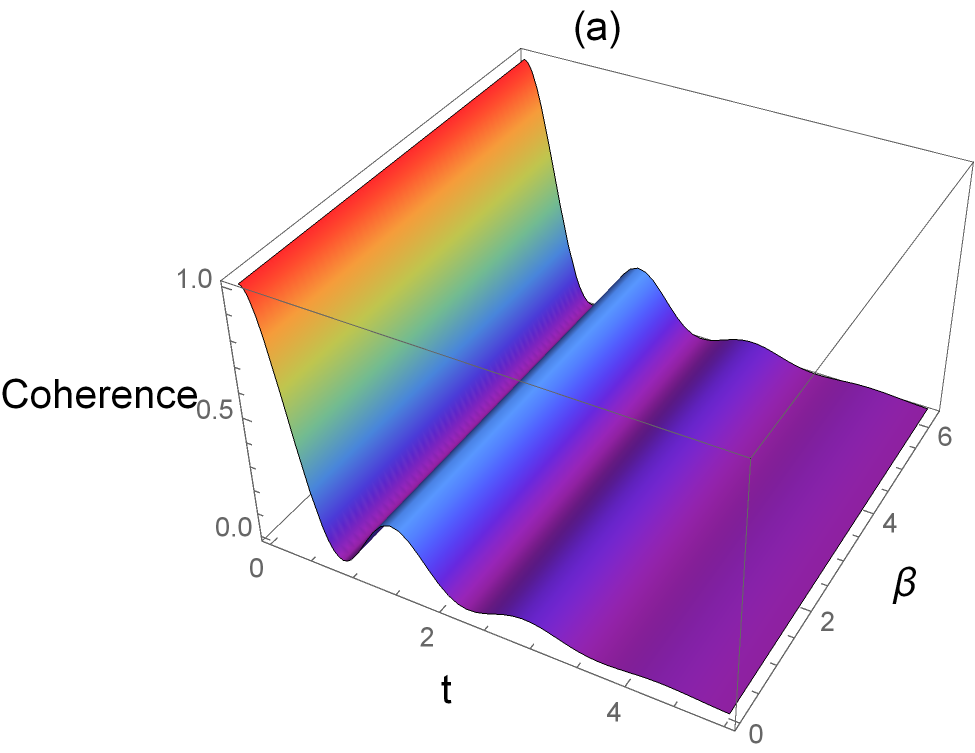}
	\includegraphics[width=4cm,height=3cm]{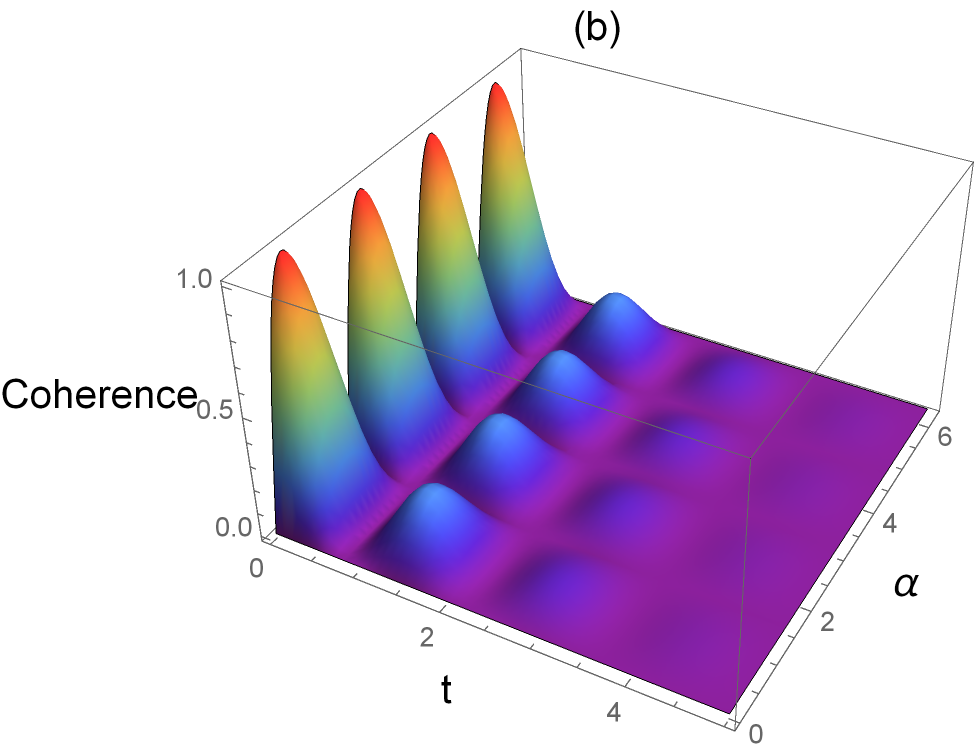}
	\caption{Three-dimensional plot of quantum coherence dynamics of a V-type three-level system as a function of initial state phases in the strong coupling regime ($\gamma_0/\kappa=10$, $\gamma_0=1$) . The initial state is in $|\psi(0)\rangle =\left(\cos\alpha|A\rangle+e^{i\beta}\sin\alpha|B\rangle\right)_{S} \otimes|0\rangle_{E}$. (a) $\alpha=\frac{\pi}{4}$ and (b) $\beta=0$. The other parameters are $\Delta=0$, $p=0$, $p_r=0$, and $\theta=0$.}
	\label{figure:2}
\end{figure}

\subsection{Atom-cavity resonance ($\Delta$ = 0)}
Secondly, when the atomic transition frequency is resonant with the cavity eigenfrequency, we investigate effects of SGI parameter $\theta$, weak measurement strength $p$ and reversing measurement strength $p_r$ on the quantum coherence dynamics of the V-type atom in the weak ($\gamma_0/\kappa=0.1$) and strong ($\gamma_0/\kappa=10$) coupling regimes, respectively.

In FIG. 3, we analyse the effect of SGI parameter $\theta$ on the quantum coherence of different initial states when $p=0$, $p_r=0$ and $\Delta=0$. FIG. 3(a) and FIG. 3(b) describe the quantum coherence dynamics of an initially maximal coherent state in the weak and strong regimes, respectively. From FIG. 3(a) (in the weak regime), we find that the quantum coherence always monotonously decays and asymptotically reaches 0 from 1.0 for all $\theta$ values. However, the decay rate depends on the $\theta$ values. Specifically, it will increase as $\theta$ rises. Compared with FIG. 3(a), the significant difference in FIG. 3(b) (in the strong regime) is that, due to the feedback and memory effects of the environment, the quantum coherence will very rapidly reduce to zero, known as coherence sudden death (CSD), and then suddenly increase to a peak value, known as coherence sudden birth (CSB). Eventually, it approaches zero after several immediate coherence death-birth transitions. For example, when $\theta=1.0$, the quantum coherence first rapidly reduces to zero and then suddenly increases to a peak value of 0.38. After three immediate coherence death-birth transitions, it eventually approaches zero. Besides, the bigger the SGI parameter is, the larger the peak value is and the shorter the revival time is. Therefore, for the initially maximal coherent state, the strong coupling between the cavity and the environment can induce several immediate coherence sudden death-birth transitions, and the SGI parameter has a negative effect on protecting the quantum coherence.

FIG. 3(c) and FIG. 3(d) show the evolutions of an initially partial coherent state with the coherence 0.87 in the weak and strong regimes, respectively. From FIG. 3(c), we can see that, in the weak regime, the quantum coherence gradually decays to zero from 0.87 and its decay rate will increase as $\theta$ rises. Compared to FIG. 3(a), the quantum coherence in FIG. 3(c) will undergo a sudden death-birth transition and then tend to a stable value as $\theta>0.6$, for instance, this steady value is equal to 0.06 when $\theta=1$ (red solid line). In the strong coupling regime, as shown in FIG. 3(d), there are several immediate coherence sudden death-birth transitions, and the bigger SGI parameter corresponds to the larger peak value and the shorter revival time, which are different from FIG. 3(c) but similar to FIG. 3(b). The steady value is also equal to 0.06 when $\theta=1$, which is the same as FIG. 3(c). Thus for the initially partial coherent state, the larger SGI parameter ($\theta>0.6$) can induce quantum coherence sudden death-birth transitions though it has also a negative effect on protecting the quantum coherence.

FIG. 3(e) and FIG. 3(f) display the evolutions of an initially incoherent state in the weak and strong regimes, respectively. We find that the non-zero SGI parameter will induce CSB, which is different from the cases of the initially maximal or partial coherent states. From FIG. 3(e), we can observe that, in the weak coupling regime, the quantum coherence can also be suddenly generated when $\theta\neq0$. Also, the larger the SGI parameter is, the more the coherence generated is. Besides, the coherence generated can approach 0 when $\theta<1$ but it tends to the steady value 0.5 only when $\theta=1$. However, in the strong coupling regime, FIG. 3(f) shows that the coherence generated undergoes several immediate coherence sudden death-birth transitions and eventually approaches 0 when $0<\theta<1$ but it oscillates to the stable value 0.5 when $\theta=1$. That is to say, the non-zero SGI parameter can effectively induce CSB in the incoherent state.
\begin{figure}[tbp]
	\includegraphics[width=4cm,height=3cm]{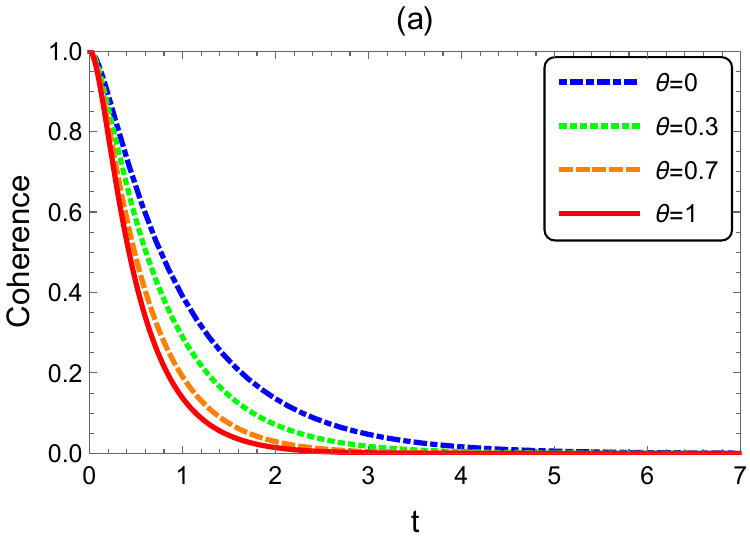}
	\includegraphics[width=4cm,height=3cm]{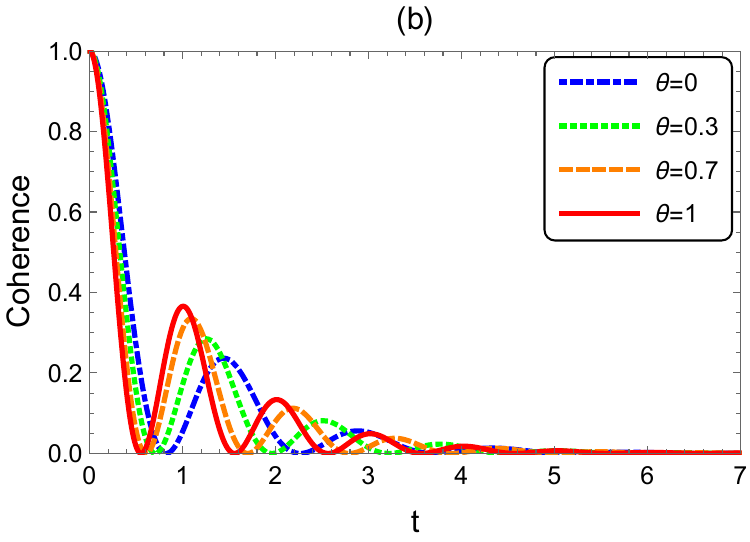}
    \includegraphics[width=4cm,height=3cm]{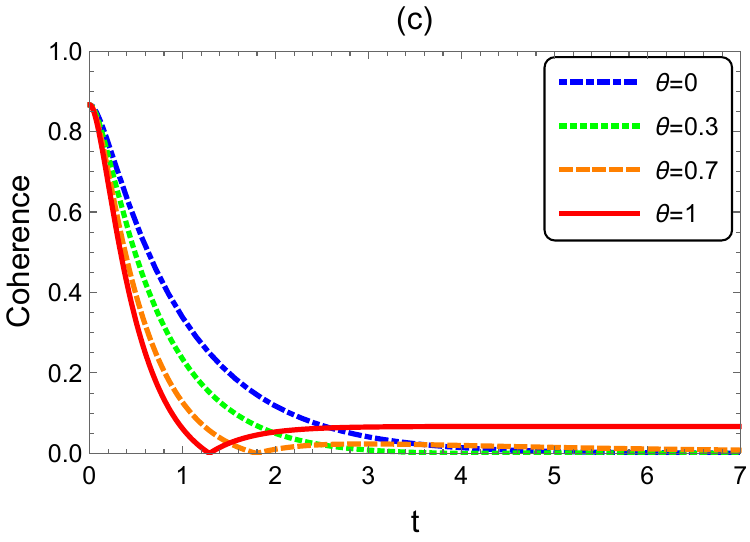}
    \includegraphics[width=4cm,height=3cm]{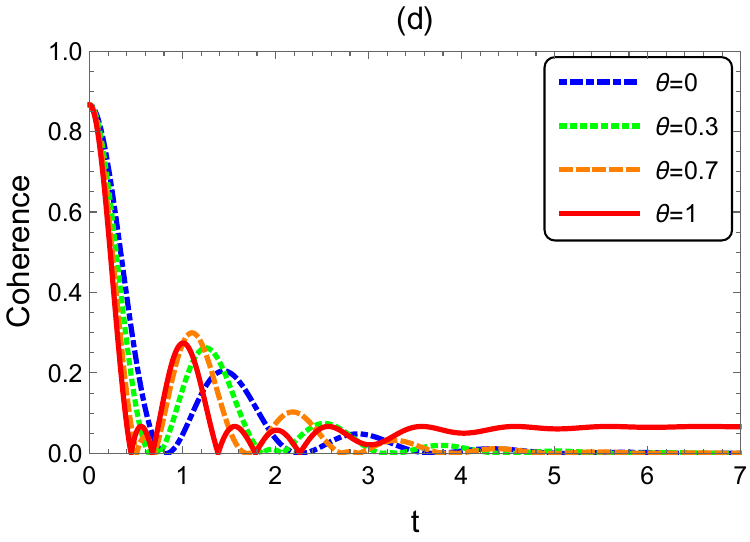}
    \includegraphics[width=4cm,height=3cm]{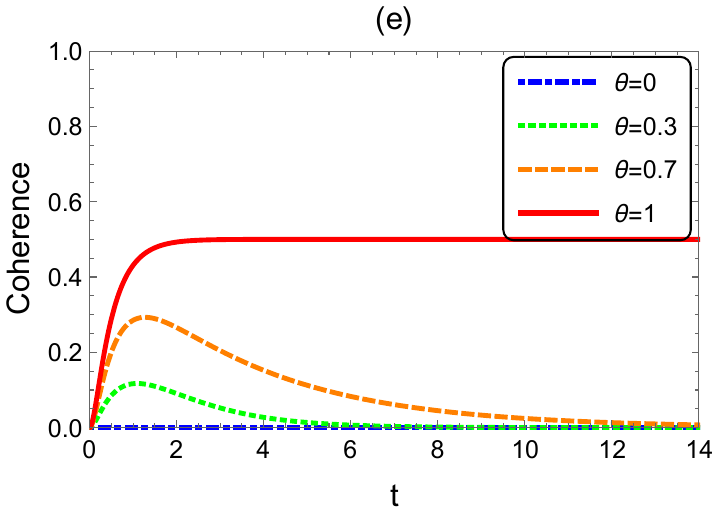}
    \includegraphics[width=4cm,height=3cm]{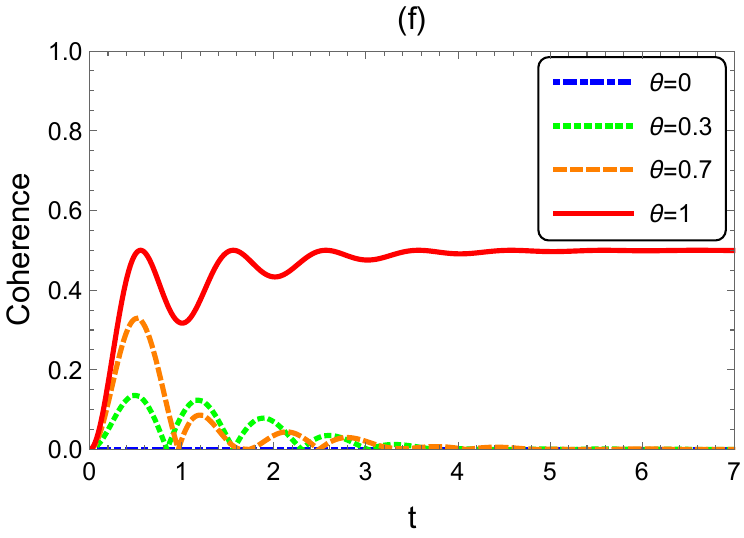}
	\caption{Quantum coherence dynamics of a V-type three-level system with $\theta=0$ (blue dot dashed line), $\theta=0.3$ (green dotted line), $\theta=0.7$ (orange dashed line), and $\theta=1$ (red solid line). The initial states are defined as follows: $|\psi(0)\rangle=\frac{\sqrt{2}}{2}(|A\rangle+|B\rangle)_{S} \otimes|0\rangle_{E}$ for (a)(b), $|\psi(0)\rangle=(\frac{1}{2}|A\rangle+\frac{\sqrt{3}}{2}|B\rangle)_{S} \otimes|0\rangle_{E}$ for (c)(d), and $|\psi(0)\rangle=|B\rangle_{S} \otimes|0\rangle_{E}$ for (e)(f). (a)(c)(e) $\gamma_0/\kappa=0.1$, $\kappa=1$ (in the weak coupling regime), (b)(d)(f) $\gamma_0/\kappa=10$, $\gamma_0=1$ (in the strong coupling regime). And the other parameters are $\Delta=0$, $p=0$, and $p_r=0$.}
	\label{figure:3}
\end{figure}

\begin{figure}[tbp]
	\includegraphics[width=4cm,height=3cm]{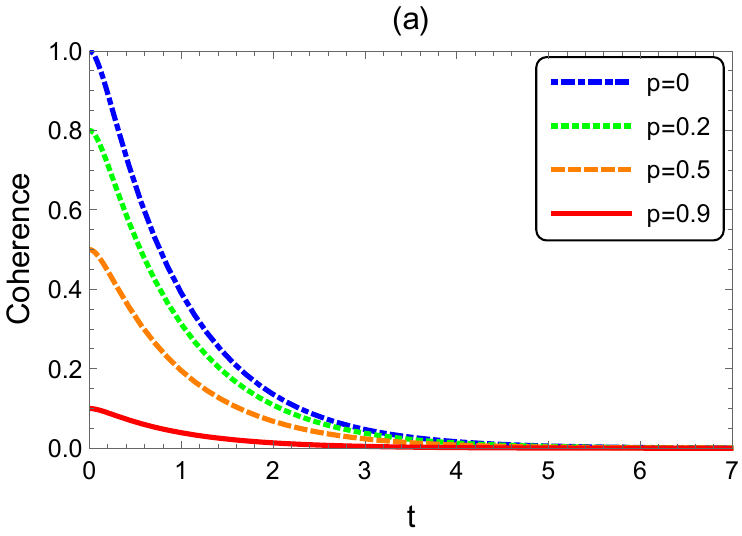}
	\includegraphics[width=4cm,height=3cm]{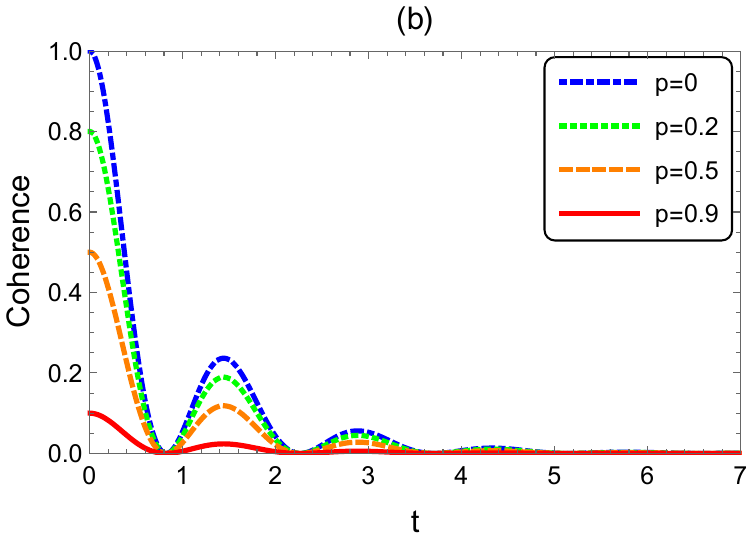}
    \includegraphics[width=4cm,height=3cm]{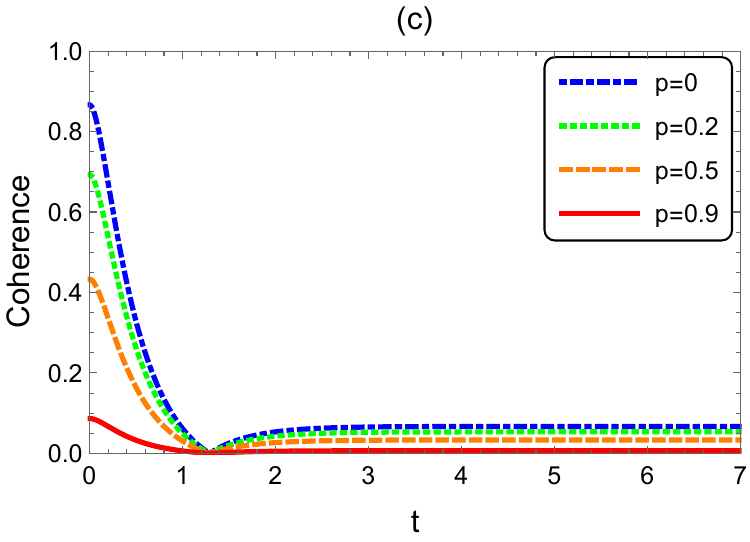}
    \includegraphics[width=4cm,height=3cm]{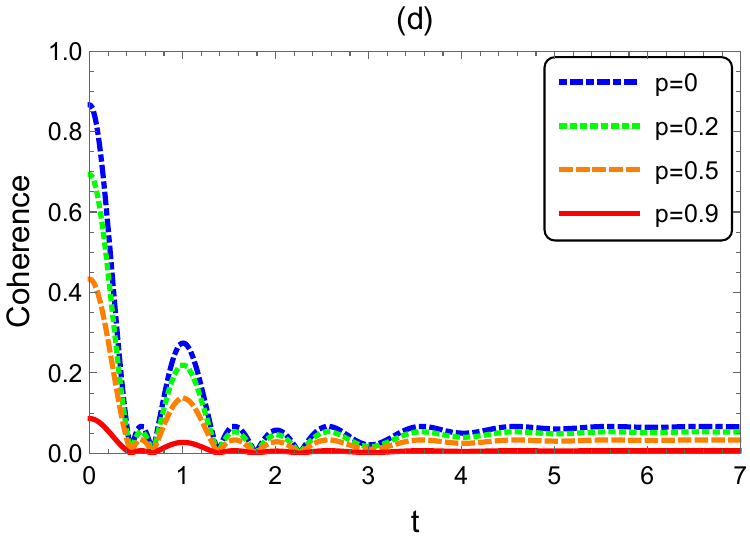}
    \includegraphics[width=4cm,height=3cm]{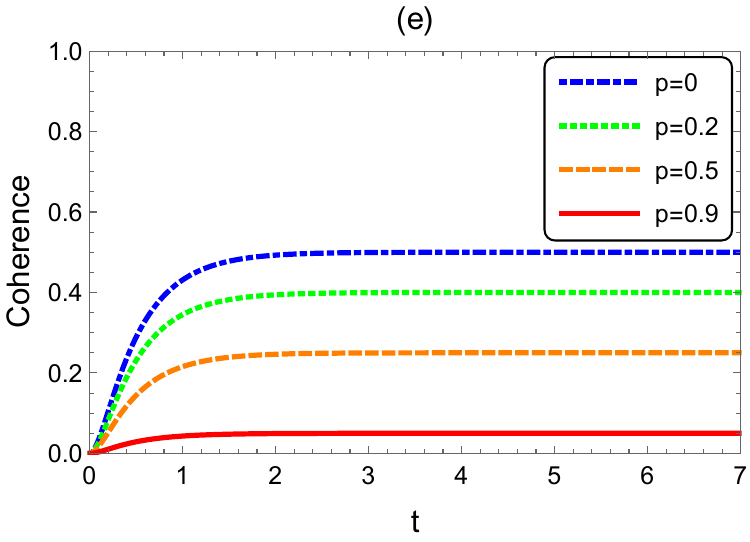}
    \includegraphics[width=4cm,height=3cm]{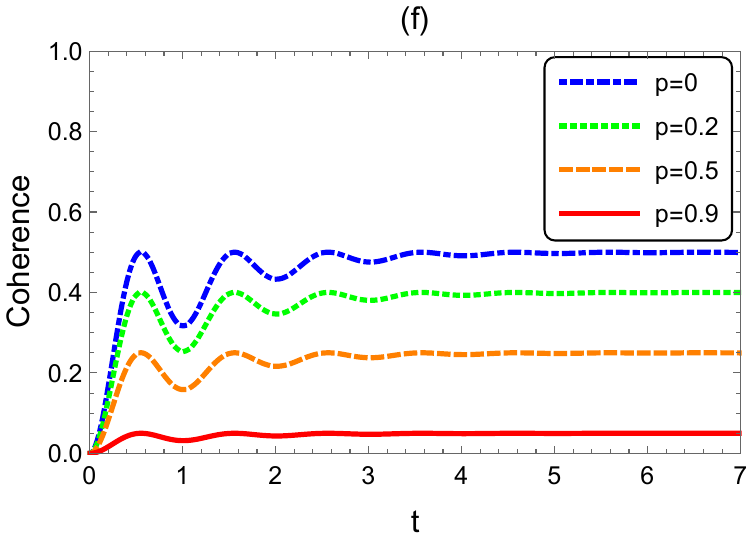}
	\caption{Quantum coherence dynamics of a V-type three-level system with $p=0$ (blue dot dashed line), $p=0.2$ (green dotted line), $p=0.5$ (orange dashed line), and $p=0.9$ (red solid line). The initial states are defined as follows: $|\psi(0)\rangle=\frac{\sqrt{2}}{2}(|A\rangle+|B\rangle)_{S} \otimes|0\rangle_{E}$ for (a)(b), $|\psi(0)\rangle=(\frac{1}{2}|A\rangle+\frac{\sqrt{3}}{2}|B\rangle)_{S} \otimes|0\rangle_{E}$ for (c)(d), and $|\psi(0)\rangle=|B\rangle_{S} \otimes|0\rangle_{E}$ for (e)(f). (a)(c)(e) $\gamma_0/\kappa=0.1$, $\kappa=1$ (in the weak coupling regime), (b)(d)(f) $\gamma_0/\kappa=10$, $\gamma_0=1$ (in the strong coupling regime). (a)(b) $\theta=0$, (c)(d)(e)(f) $\theta=1$. And the other parameters are $\Delta=0$ and $p_r=0$.}
	\label{figure:4}
\end{figure}
In FIG. 4, we describe the evolutionary dynamics of different initial states under weak measurement strengths $p$ when the SGI parameter is optimal. FIG. 4(a) and FIG. 4(b) show the quantum coherence of an initially maximal coherent state with the coefficients $D_{A}(0)=D_{B}(0)=\frac{\sqrt{2}}{2}$ and $\theta=0$. FIG. 4(a) (in the weak coupling regime) indicates that, as $p$ increases, the initial coherence will become small and gradually decay to zero. FIG. 4(b) (in the strong coupling regime) indicates that the coherence approaches zero after several immediate coherence death-birth transitions, and the peak increases as $p$ decreases, which is different from FIG. 4(a). Thus the weak measurement is a negative effect on protecting the quantum coherence.

FIG. 4(c) and FIG. 4(d) display the evolutions of an initially partial coherent state when $\theta=1$. From FIG. 4(c) (in the weak coupling regime), we can see that the coherence does not gradually decay to zero, but rather tends to a steady value after undergoing an immediate coherence death-birth transition. This behavior is different from that observed in FIG. 4(a). The quantum coherence in FIG. 4(d) (in the strong coupling regime) tends to a steady value after undergoing several immediate coherence death-birth transitions, which is different from FIG. 4(c). The steady value will increase with $p$ decreasing, which is consistent with FIG. 4(c). The peak will also increase as $p$ decreases, which is similar to FIG. 4(b). Thus the weak measurement is also a negative effect on protecting the quantum coherence.

FIG. 4(e) and FIG. 4(f) exhibit the quantum coherence dynamics when the initial atomic state is incoherent and $\theta=1$. We know that in the weak coupling regime (FIG. 4(e)), the coherence can be suddenly generated and monotonically tend to a steady value. The coherence can be generated more quickly, and the steady value is larger as $p$ decreases. In FIG. 4(f), the coherence will oscillate towards a stable value after it is suddenly generated, which is different from the behavior observed in FIG. 4(e). For the same $p$ value, the stable value in FIG. 3(f) is the same as that in FIG. 4(e). Thus the weak measurement has also a negative effect on CSB.
\begin{figure}[tbp]
	\includegraphics[width=4cm,height=3cm]{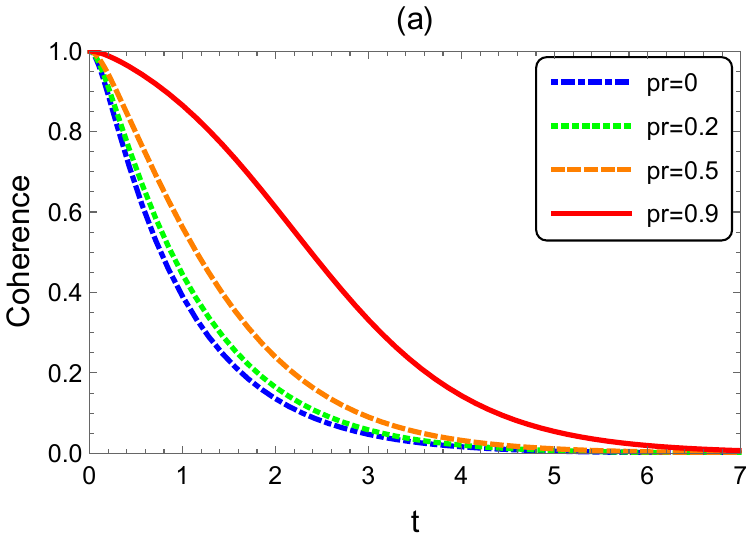}
	\includegraphics[width=4cm,height=3cm]{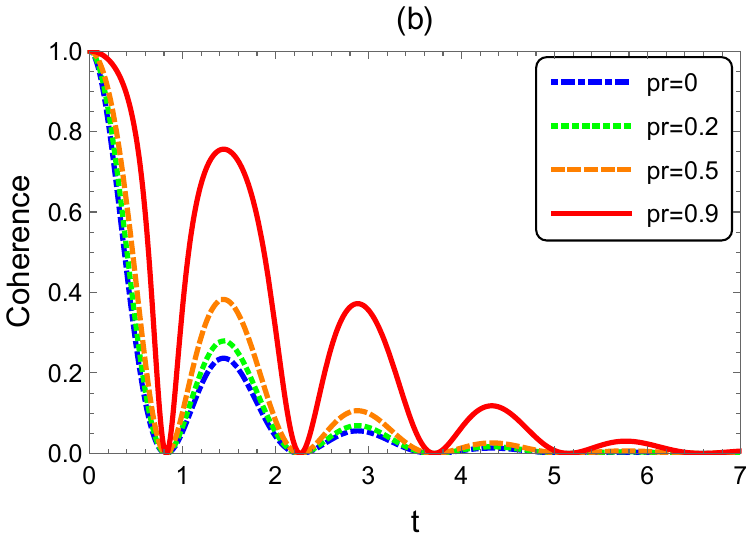}
    \includegraphics[width=4cm,height=3cm]{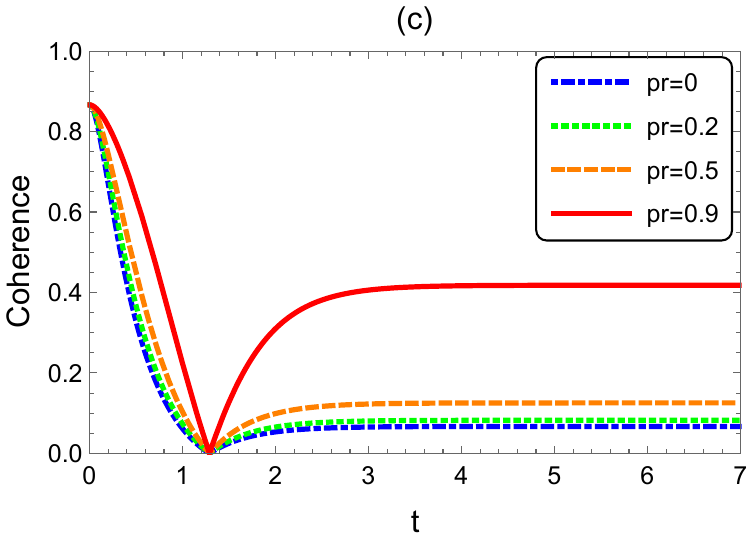}
    \includegraphics[width=4cm,height=3cm]{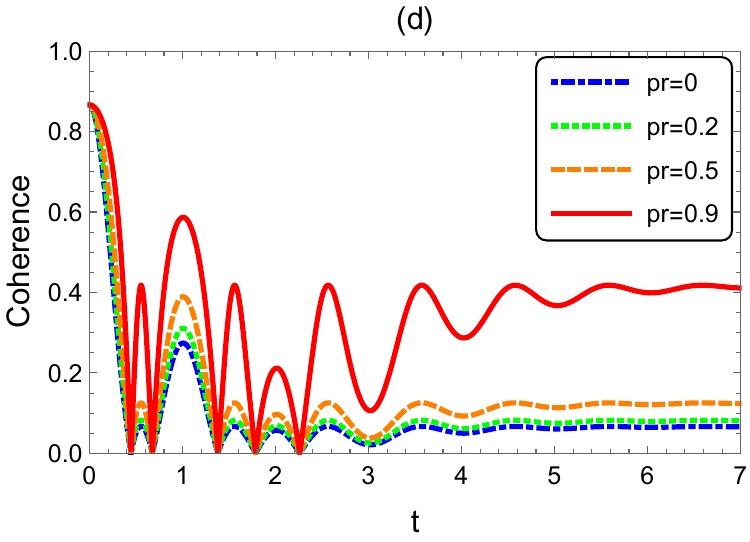}
    \includegraphics[width=4cm,height=3cm]{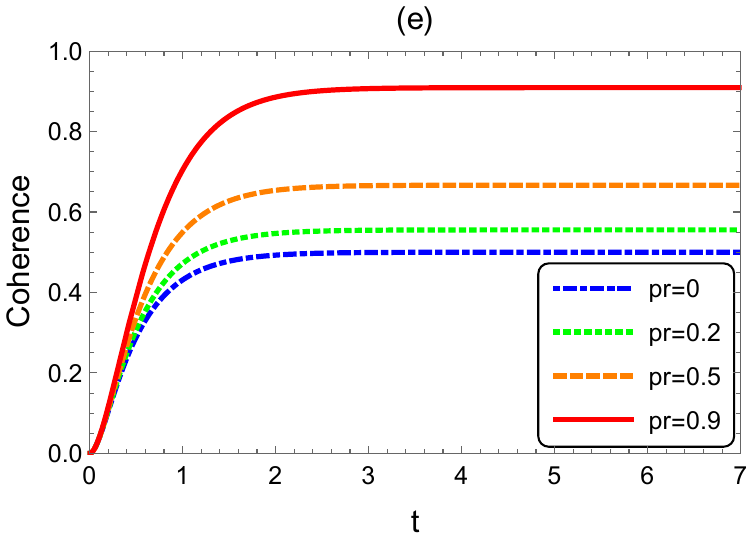}
    \includegraphics[width=4cm,height=3cm]{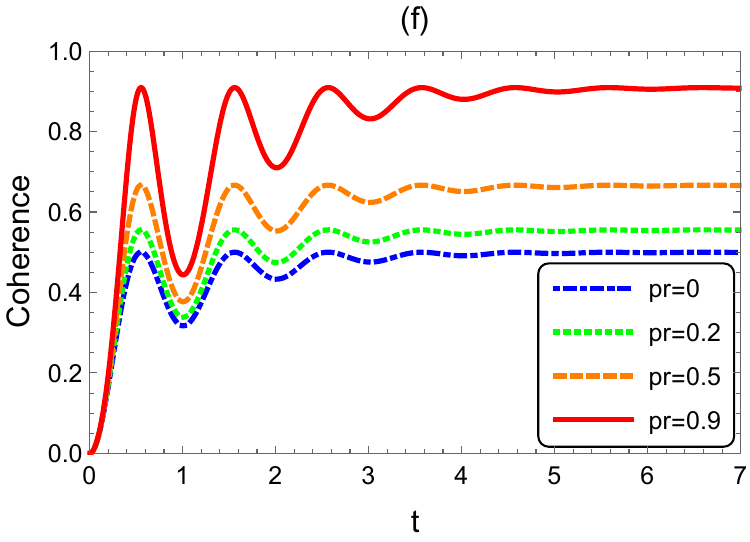}
	\caption{Quantum coherence dynamics of a V-type three-level system with $p_r=0$ (blue dot dashed line), $p_r=0.2$ (green dotted line), $p_r=0.5$ (orange dashed line), and $p_r=0.9$ (red solid line). The initial states are defined as follows: $|\psi(0)\rangle=\frac{\sqrt{2}}{2}(|A\rangle+|B\rangle)_{S} \otimes|0\rangle_{E}$ for (a)(b), $|\psi(0)\rangle=(\frac{1}{2}|A\rangle+\frac{\sqrt{3}}{2}|B\rangle)_{S} \otimes|0\rangle_{E}$ for (c)(d), and $|\psi(0)\rangle=|B\rangle_{S} \otimes|0\rangle_{E}$ for (e)(f). (a)(c)(e) $\gamma_0/\kappa=0.1$, $\kappa=1$ (in the weak coupling regime), (b)(d)(f) $\gamma_0/\kappa=10$, $\gamma_0=1$ (in the strong coupling regime). (a)(b) $\theta=0$, (c)(d)(e)(f) $\theta=1$. And the other parameters are $\Delta=0$ and $p=0$.}
	\label{figure:5}
\end{figure}

In FIG. 5, we analyze the effects of reversing measurement strength $p_r$ on the quantum coherence in different coupling regimes and initial states when $p=0$ and the SGI parameter is optimal. FIG. 5(a) and FIG. 5(b) show the evolutions of an initially maximal coherent state when $\theta=0$ under the weak measurement reversal. From FIG. 5(a) (in the weak coupling regime), we can see that the coherences monotonically decrease and asymptotically approach 0 from 1.0, regardless of the $p_r$ values. Additionally, the larger the $p_r$ is, the slower the coherence decays. For the strong coupling (FIG. 5(b)), the coherences decay from 1 and then undergo several immediate coherence sudden death-birth transitions, finally tending to 0. Also, the coherence peak increases as the reversing measurement strength increases, and the peaks in FIG.5(b) are obviously bigger than those in FIG.3(b). This indicates that the weak measurement reversal can enhance CSB. Thus the weak measurement reversal has a positive effect on protecting and generating the quantum coherence.

FIG. 5(c) and FIG. 5(d) exhibit the evolutions of an initially partial coherent state with the coherence 0.87 under the weak measurement reversal, and we set the optimal SGI parameter $\theta=0$. In the weak coupling regime (FIG. 5(c)), we know that the coherence rapidly and monotonically decreases and undergoes one sudden death-birth transition, and then tends to a steady value. For example, this steady value can reach 0.42 when $p_r=0.9$ (red solid line). Also, it is evident that as $p_r$ increases, the decay rate will slow down but the coherence regenerated will be more. Similar to FIG. 5(c), the weak measurement reversal can also suppress its decay and promote its generation for quantum coherence in FIG. 5(d), and the stable value depends on the $p_r$ value. Thus the weak measurement reversal has also a positive effect on protecting and generating the quantum coherence.

FIG. 5(e) and FIG. 5(f) show that the evolutions of an initially incoherent state under the weak measurement reversal. From the blue dot dashed line ($p_r=0$) in FIG. 5(e) (the weak coupling regime), it can be seen that the coherence is suddenly generated and monotonically increases a stable value 0.5. Additionally, as $p_r$ increases, a larger stable value can be obtained, for example, the steady value is equal to 0.9 when $p_r=0.9$. In the strong coupling regime (FIG. 5(f)), the coherence will oscillate towards a stable value after it is suddenly generated, which is different from FIG. 5(e). For the same $p_r$ value, the stable value in FIG. 5(f) is the same as that in FIG. 5(e). As a result, the weak measurement reversal can very effectively protect and generate the quantum coherence.

\subsection{Atom-cavity detuning ($\Delta\neq 0$)}
Next, we focus on the effects of detuning $\Delta$ on quantum coherence of different initial states  in the weak ($\gamma_0/\kappa=0.1$) and strong ($\gamma_0/\kappa=10$) coupling regimes when $p=0$, $p_r=0.9$ and the SGI parameter is optimal.

\begin{figure}[tbp]
	\includegraphics[width=4cm,height=3cm]{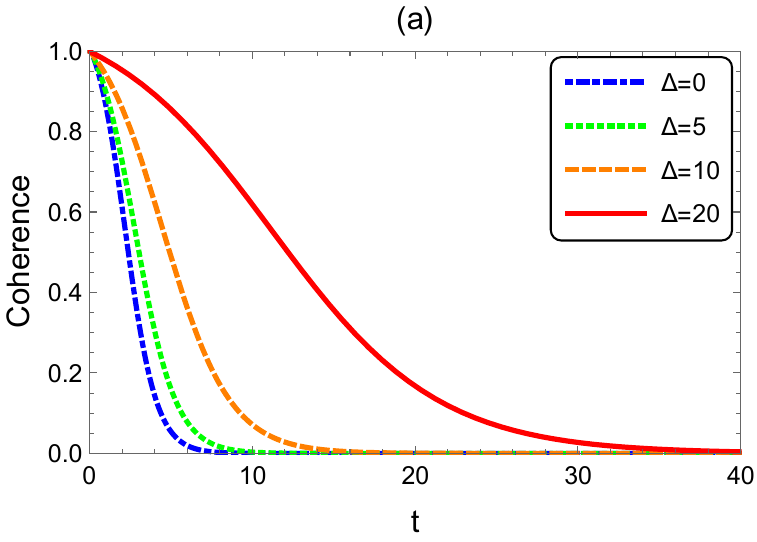}
	\includegraphics[width=4cm,height=3cm]{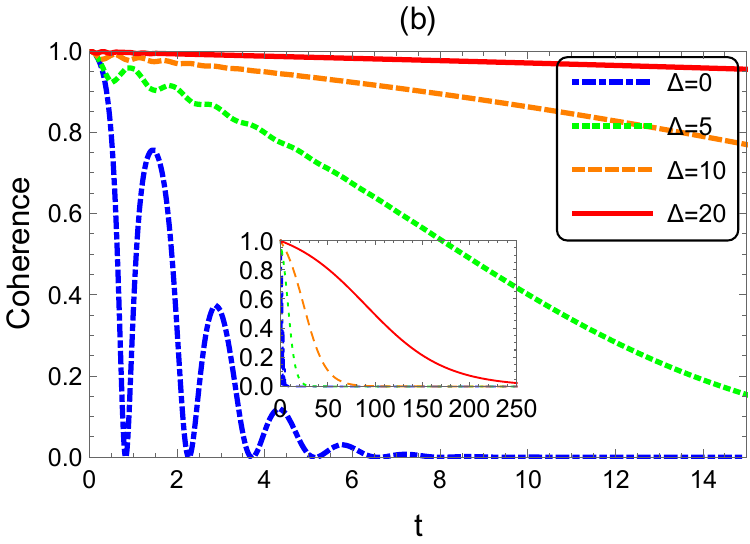}
    \includegraphics[width=4cm,height=3cm]{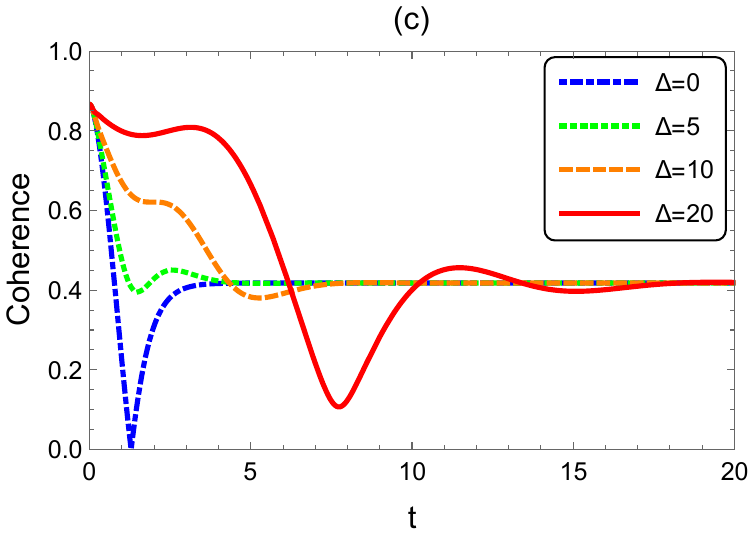}
    \includegraphics[width=4cm,height=3cm]{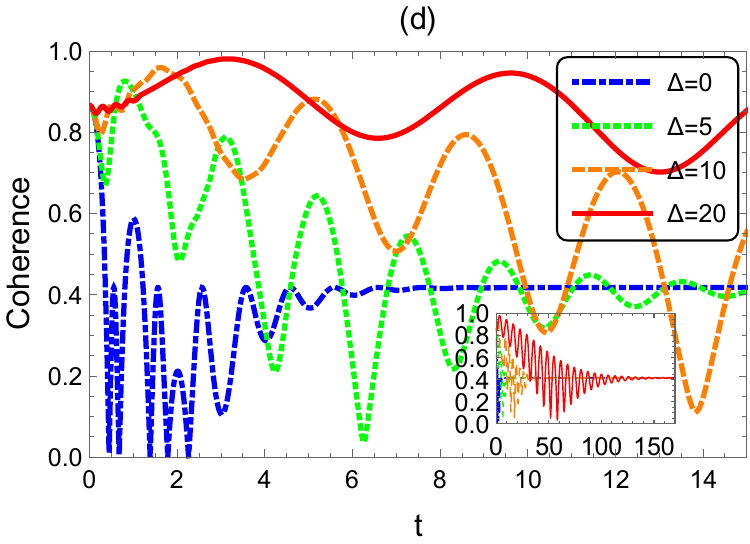}
    \includegraphics[width=4cm,height=3cm]{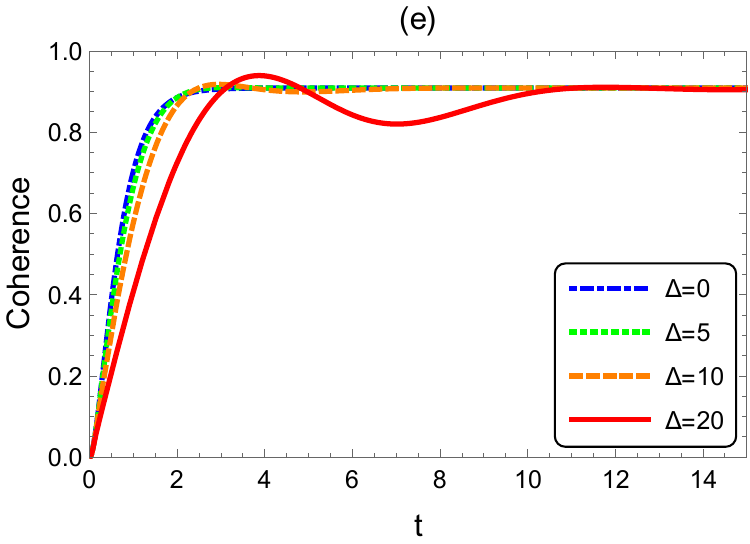}
    \includegraphics[width=4cm,height=3cm]{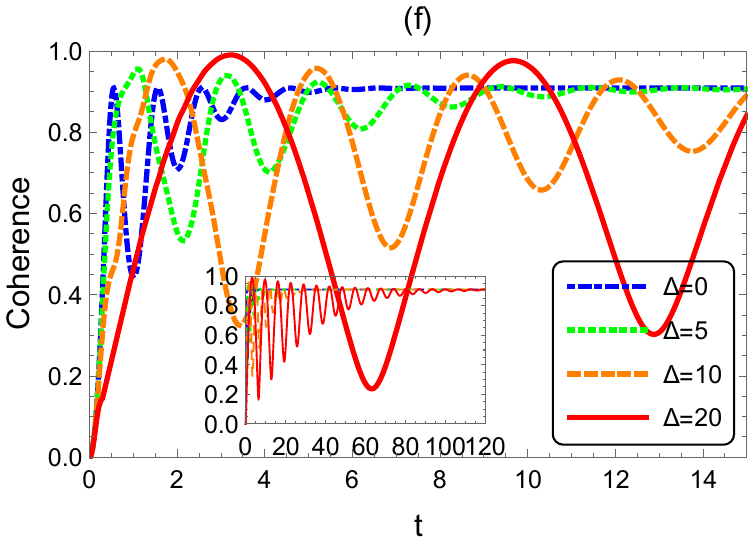}
	\caption{Quantum coherence dynamics of a V-type three-level system with $\Delta=0$ (blue dot dashed line), $\Delta=5$ (green dotted line), $\Delta=10$ (orange dashed line), and $\Delta=20$ (red solid line). The initial states are defined as follows: $|\psi(0)\rangle=\frac{\sqrt{2}}{2}(|A\rangle+|B\rangle)_{S} \otimes|0\rangle_{E}$ for (a)(b), $|\psi(0)\rangle=(\frac{1}{2}|A\rangle+\frac{\sqrt{3}}{2}|B\rangle)_{S} \otimes|0\rangle_{E}$ for (c)(d), and $|\psi(0)\rangle=|B\rangle_{S} \otimes|0\rangle_{E}$ for (e)(f). (a)(c)(e) $\gamma_0/\kappa=0.1$, $\kappa=1$ (in the weak coupling regime), (b)(d)(f) $\gamma_0/\kappa=10$, $\gamma_0=1$ (in the strong coupling regime). (a)(b) $\theta=0$, (c)(d)(e)(f) $\theta=1$. And the other parameters are $p=0$ and $p_r=0.9$.}
	\label{figure:6}
\end{figure}

In FIG. 6(a) and FIG. 6(b), we plot the curves of coherence dynamics of an initially maximal coherent state when $\theta=0$. FIG. 6(a) (in the weak coupling regime) shows that the quantum coherence will monotonically reduce and asymptotically reaches zero from 1.0 and its decay rate will become small with $\Delta$ increasing. From the blue dot dashed line ($\Delta=0$) in FIG. 6(b), which is in the strong coupling regime, we observe that the quantum coherence eventually approaches zero after undergoing several immediate coherence death-birth transitions. However, if $\Delta>0$, CSD and CSB will not occur, that is, the detuning will greatly improve the robustness of quantum coherence. Specifically, when $\Delta=20$ and $\gamma_0/\kappa=10$, the coherence has always been very close to 1.0 as time progresses (red solid line), namely, the maximal coherent state can be very effectively protected. Therefore, the detuning can not only avoid CSD but also protect very effectively the maximal coherent state in the strong coupling regime.

To demonstrate how detuning affects an initially partial coherent state with the coherence 0.87, we give FIG. 6(c) and FIG. 6(d) for different coupling regimes when $\theta=1$. FIG. 6(c) (in the weak coupling regime) indicates that there is an immediate coherence death-birth transition only when $\Delta=0$ (blue dot dashed line) while the CSD and the CSB will not occur when $\Delta>0$. As the detuning increases, the decay rate will obviously become smaller, but the coherence will tend to the same steady value of 0.42. In the strong coupling regime (FIG. 6(d)), there are several immediate coherence death-birth transitions only when $\Delta=0$. As the detuning increases, both the amplitude and frequency will enlarge and the peaks may be bigger than 0.87, but there is the same steady value, which is the same as FIG. 6(c). Specifically, the quantum coherence will rise to the maximum 0.98 at $t=3.2$ from 0.87 and then oscillate to the steady value 0.42 when $\Delta=20$ and $\gamma_0/\kappa=10$. Namely, the system evolves to a coherent state with $\xi(\rho)=0.98$ and then approaches a stable coherent state with $\xi(\rho)=0.42$. Substituting the relevant parameters in FIG.6(d) into Eq.~(\ref{Eq29}), the atomic density operators can be respectively expressed as
\begin{equation}\label{Eq31}
\begin{aligned}
& \rho_{t=3.2}^{0.98}=
& \left(\begin{array}{ccc}
0.0196 & 0 & 0 \\
0 & 0.4938 & 0.4114-0.2665i \\
0 & 0.4114+0.2665i & 0.4866
\end{array}\right)
\end{aligned}
\end{equation}
and
\begin{equation}\label{Eq32}
\begin{aligned}
& \rho_{stable}^{0.42}=
& \left(\begin{array}{ccc}
0.582 & 0 & 0 \\
0 & 0.209 & -0.209 \\
0 & -0.209 & 0.209
\end{array}\right)
\end{aligned}
\end{equation}
Therefore, for an initially partial coherent state, the detuning may not only avoid CSD but also prepare the coherent states.

FIG. 6(e) and FIG. 6(f) respectively illustrate the evolutions of an initially incoherent state in the weak and strong coupling regimes when $\theta=1$. In FIG. 6(e), we can observe from the blue dot dashed line that the quantum coherence can be quickly generated to a steady value 0.91 from 0 when the detuning $\Delta=0$. The detuning has no significant effect on the coherence dynamics in the weak coupling regime. From FIG. 6(f), we can observe that the detuning has very significant effect on the coherence dynamics in the strong coupling regime, i.e. the coherence increases to a peak from 0 and then oscillates to the steady value 0.91, and both the peak and the period of coherence will become larger with the detuning increasing. For example, the peak value is close to 0.99 at $t=3.2$ when $\Delta=20$ and $\gamma_0/\kappa=10$. Namely, the system evolves to a coherent state with $\xi(\rho)=0.99$ and then approaches a stable coherent state with $\xi(\rho)=0.91$. Substituting the relevant parameters in FIG.6(f) into Eq.~(\ref{Eq29}), the atomic density operators can be respectively expressed as
\begin{equation}\label{Eq33}
\begin{aligned}
& \rho_{t=3.2}^{0.99}=
& \left(\begin{array}{ccc}
0.010 & 0 & 0 \\
0 & 0.502 & -0.049-0.493i \\
0 & -0.049+0.493i & 0.488
\end{array}\right)
\end{aligned}
\end{equation}
and
\begin{equation}\label{Eq34}
\begin{aligned}
& \rho_{stable}^{0.91}=
& \left(\begin{array}{ccc}
0.091 & 0 & 0 \\
0 & 0.4545 & -0.4545 \\
0 & -0.4545 & 0.4545
\end{array}\right)
\end{aligned}
\end{equation}
Hence, for the incoherent state, the detuning can be very effective in promoting CSB and preparing the coherent states.

\subsection{Physical explanation}
In the following, the physical explanation of the results above is given. Firstly, we explain the effect of SGI parameter on quantum coherence dynamics. $\theta=1$ means the two dipole moments are parallel, which suggests that the exciton transition rate between the atomic energy levels is at its maximum. Therefore, the bigger the SGI parameter is, the faster the coherence decays (refer to FIG. 3(a)(b)(c)(d)) and the more the coherence is generated (refer to FIG. 3(e)(f)). Secondly, we analyze the effects of cavity-environment coupling on quantum coherence dynamics, including both weak and strong coupling cases. In the weak coupling regime, the quantum coherence will decrease monotonously due to the dissipation of quantum information into the reservoir, as shown in FIG. 3(a)(c)(e). In the strong coupling regime, the feedback and memory effects of the environment will cause the quantum coherence to exhibit damped oscillations, as shown in FIG. 3(b)(d)(f). Thirdly, we separately discuss the effects of weak measurement and weak measurement reversal on quantum coherence dynamics. When the atom is subjected to a prior weak measurement, it will collapse to the ground state, thereby destroying the initial quantum coherence. The amount of quantum coherence destroyed increases with the strength of the weak measurement, which is consistent with the results in FIG. 4. If the atom is subjected to a post weak measurement reversal, it will flip from the ground state to the excited state with a certain probability, thus effectively protecting the quantum coherence. The probability of flipping increases with the strength of the reversing measurement, which is consistent with the results in FIG. 5. Finally, we consider the effect of detuning on quantum coherence dynamics. Under detuning, quantum information can be effectively trapped in the atom-cavity system, thus significantly reducing the information dissipated through the environment. The larger the detuning is, the smaller the dissipated information is, and the better the coherence is protected, as shown in FIG. 6.

\section{Conclusion}
In summary, we have investigated the quantum coherence dynamics of a V-type atom in a dissipative cavity under detuning and weak measurement reversal. Firstly, we analytically obtained the expression for the atomic quantum coherence using the $\ell_{1}$ norm. Secondly, we examined the influence of the SGI parameter, cavity-environment coupling, detuning between the atom and the cavity, and weak measurement and its reversal on the quantum coherence dynamics. We found that, the strong coupling can induce the CSD and the CSB, while the non-zero SGI parameter only induces the CSB. Detuning, however, may avoid both the CSD and the CSB. The coherence always monotonously decays and asymptotically reaches 0 from 1.0 under the weak coupling, while it undergoes several immediate coherence sudden death-birth transitions and then approaches 0 under the strong coupling. Moreover, detuning and weak measurement reversal can very effectively protect quantum coherence, while the SGI parameter, weak measurement, and strong coupling can accelerate its attenuation. The SGI parameter, detuning, weak measurement reversal, and strong coupling all promote the generation of coherence, whereas weak measurement alone can suppress it. In particular, the maximal coherent state can be very effectively protected and the coherent state can be prepared if all parameters are selected appropriately. For example, the stable coherence generated can reach 0.91 when $\theta$=1.0, $p$=0 and $p_r$=0.9 regardless of whether it is in the weak or strong coupling regime. Under the detuning, the maximal coherence state can be very effectively protected in the strong coupling regime when $\theta$=1.0, $p$=0, $p_r$=0.9 and $\Delta$=20. For the incoherent state, the coherence peak generated increases with rising detuning, and the maximal coherence approaches 0.99 when $\theta$=1.0, $p$=0, $p_r$=0.9 and $\Delta$=20. We also provided physical interpretations for all these results.

\begin{acknowledgments}
This work was supported by the Foundation Xiangjiang-Laboratory (XJ2302001), Changsha, Hunan, China.
\end{acknowledgments}

\begin{widetext}

\end{widetext}

\end{document}